\documentclass[aps,pop,preprint]{revtex4-1}
\usepackage{amsmath}
\usepackage{amssymb}
\usepackage{graphicx}

\begin{document}
\author{A.E.~Fraser}
\affiliation{University of Wisconsin-Madison, Madison, Wisconsin 53706, U.S.A.}
\author{P.W.~Terry}
\affiliation{University of Wisconsin-Madison, Madison, Wisconsin 53706, U.S.A.}
\author{E.G.~Zweibel}
\affiliation{University of Wisconsin-Madison, Madison, Wisconsin 53706, U.S.A.}
\author{M.J.~Pueschel}
\affiliation{University of Wisconsin-Madison, Madison, Wisconsin 53706, U.S.A.}
\title{Coupling of Damped and Growing Modes in Unstable Shear Flow}
\begin{abstract}
	Analysis of the saturation of the Kelvin-Helmholtz (KH) instability is undertaken to determine the extent to which the conjugate linearly stable mode plays a role. For a piecewise-continuous mean flow profile with constant shear in a fixed layer, it is shown that the stable mode is nonlinearly excited, providing an injection-scale sink of the fluctuation energy similar to what has been found for gyroradius-scale drift-wave turbulence. Quantitative evaluation of the contribution of the stable mode to the energy balance at the onset of saturation shows that nonlinear energy transfer to the stable mode is as significant as energy transfer to small scales in balancing energy injected into the spectrum by the instability. The effect of the stable mode on momentum transport is quantified by expressing the Reynolds stress in terms of stable and unstable mode amplitudes at saturation, from which it is found that the stable mode can produce a sizable reduction in the momentum flux.
\end{abstract}

\maketitle

\section{Introduction}
Shear flows are encountered in a variety of different systems. In the atmosphere, shear-flow instabilities are observed in cloud patterns\cite{Browning}. In fusion devices, turbulence generates shearing zonal flows whose potential for instability can significantly impact confinement\cite{Dimits, Rogers}. Shear-flow instabilities are especially important in astrophysics. There, differential velocities are produced by a host of processes in a variety of settings, including jets driven by accretion of mass onto compact objects such as protostars or supermassive black holes, intergalactic clouds falling into a galaxy, and galaxies plowing through the intracluster medium. In astrophysical systems, it is thought that shear-flow instabilities induce formation of a turbulent shear layer, resulting in entrainment of material through turbulent momentum transport\cite{Churchwell}, thermal and chemical mixing\cite{Kwak}, and the possibility of acceleration of particles to high energy\cite{Rieger}.

Shear-flow instability in a plasma with a uniform magnetic field perpendicular to both the flow and shear directions has the same growth rate as hydrodynamic shear flow with the same profile, illustrating that strong connections exist between hydrodynamic and plasma shear-flow instabilities. The number of potential applications in both systems makes quantitative models of turbulence driven by sheared flows highly desirable. Analytical models that describe spectral properties are important because both the separation between scales and Reynolds numbers found in astrophysical systems are much larger than what can typically be obtained in converged hydrodynamic and magnetohydrodynamic (MHD) simulations\cite{Palotti, Lecoanet}.

Efforts to characterize the nonlinear state of turbulent systems like those mentioned above commonly employ the growth rate and mode structure of the dominant linearly unstable eigenmode, which, after all, drives the turbulence. Examples are mixing-length estimates of transport, which for unstable systems are built on the linear growth rate and an unstable wavenumber, and the quasilinear transport approximation, which uses the cross phase of the unstable eigenmode to approximate the fluctuating correlation responsible for transport. Such approximations are straightforward to construct because they rely on well-understood linear properties of instabilities. However, as unstable systems move into the turbulent regime, there can be no saturation if fluctuations and transport are not modified from the linear state in some essential fashion. The precise nature of such modifications is not well understood. The standard assumption is that the modifications can be treated as a cascade to smaller scales that eventually become damped, in analogy to externally forced Navier-Stokes turbulence. This type of approach overlooks stable eigenmodes at the same scales as the instability, which invariably exist as other roots of the instability dispersion relation, and may modify the dynamics at the largest scales.

In gyroradius-scale instability-driven turbulence relevant to fusion devices, it has been recognized for more than a decade that stable modes are important in turbulence and should not be neglected\cite{Baver, Hatch2012}. Such modes can be represented as eigenmodes of the linearized system, and occur at the same length scale as the driving instability. Both stable modes, which have a negative linear growth rate,\cite{Terry2006} and subdominant modes, which can have a growth rate that is positive but smaller than that of the most unstable mode\cite{MJ}, are difficult to detect in initial value simulations. When perturbations are small and only the linear dynamics are considered, these modes are negligible compared to the most unstable mode. However, as the most unstable mode grows in amplitude, nonlinear three-wave interactions between it and the stable modes can pump energy into the latter, causing them to grow and have a significant impact on the turbulence. In collisionless trapped electron mode turbulence, for example, stable modes radically change the dynamics of the system, including changing the direction of particle flux\cite{Terry2006,TerryGatto}. In recent studies of plasma microturbulence in stellerators, quasilinear calculations of energy transport cannot reproduce the results of nonlinear simulations without including every subdominant unstable mode\cite{MJ}.

While it has been demonstrated that stable modes are universally excited and can have significant impacts on turbulence in the context of gyroradius-scale instabilities in fusion plasmas, their effects have not been studied in global-scale hydrodynamic or MHD instabilities. This paper presents an analysis of a hydrodynamic system with global-scale eigenmodes, demonstrating the nonlinear excitation of stable modes and quantifying their impact on the turbulence using techniques that were successful in plasma microturbulence. An important aspect of this paper is that tools developed in previous analytic calculations for homogeneous systems are extended for analysis of nonlinear excitation in the inhomogeneous environment of unstable shear flows. In previous calculations, the PDEs that describe relevant dynamical quantities were Fourier-transformed to obtain a system of ODEs describing the time-dependence of the Fourier amplitudes. The ODEs were then linearized about an unstable equilibrium to obtain a system of equations of the form $\dot{\mathbf{f}}=\mathcal{D}\mathbf{f}$, where $\mathbf{f}(\mathbf{k},t)$ is a vector describing the state of the system at wavenumber $\mathbf{k}$, and $\mathcal{D}$ is the matrix of linear coupling coefficients. The eigenvectors of  $\mathcal{D}$ are the eigenmodes of the system, and their eigenvalues are the frequencies and growth rates. The nonlinear excitation of linearly stable modes was then demonstrated by expanding the nonlinearities of the ODEs in the basis of the linear eigenmodes. With inhomogeneous systems, eigenmodes are not obtained by Fourier-transforming the PDEs and diagonalizing a matrix. Consequently, constructing an invertible transformation between dynamical quantities and linear eigenmodes, and expanding nonlinearities in terms of the eigenmodes, requires appropriate conditioning of the problem.

The paper is organized as follows. In Sec.~II we consider an unstable shear flow and discuss its unstable and stable eigenmodes. In Sec.~III we develop a mapping of the fluctuating flow onto the linear eigenmodes that allows a quantitative description of the energy transfer between the unstable and stable modes. In Sec.~IV we use the tools of previous calculations to assess the level to which stable modes are excited relative to unstable modes in saturation. In Sec.~V we consider the impact of stable modes on turbulent momentum transport. Conclusions are presented in Sec.~VI.

Though we start from equations that describe a two-dimensional, unmagnetized shear flow, this system is identical to a magnetized shear flow where the equilibrium magnetic field is uniform and in the direction perpendicular to both the flow and the gradient of the flow\cite{Chandra}. Future work will consider the case of a magnetic field in the direction of flow.

\section{Linear Modes}
We investigate a piecewise linear equilibrium flow in the $x$ direction with variation in the $z$ direction within a finite region of width $2d$, referred to as the shear layer. The equilibrium flow is $\mathbf{v}_0 = (U(z),0,0)$, where
\begin{equation*}
U(z) = 
\begin{cases}
1 & z\geq 1 \\
z & -1\leq z\leq 1 \\
-1 & z\leq -1.
\end{cases}
\end{equation*}
Here, $U = U^*/U_0$ is the flow normalized to the flow speed $U_0$ outside the layer, $(x,z) = (x^*/d, z^*/d)$ are coordinates normalized to the layer half-width $d$, and time will be normalized by $t = t^*U_0/d$.

Constant shear in a shear layer provides the simplest shear-flow instability for which the nonlinear driving of stable modes can be described analytically. The vortex sheet\cite{Chandra} is an even simpler manifestation of shear-flow instability, but the discontinuous equilibrium flow leads to a discontinuous eigenmode structure. Consequently, the eigenmode projection of the nonlinearity, which is calculated in the following section and involves a product of derivatives of the eigenmodes [see Eq.~\eqref{deltahat}], is not well-defined.

Here, flow is assumed to be 2D ($\partial/\partial y \equiv 0$), inviscid, and incompressible. It has been shown that for unmagnetized shear flows, 2D perturbations are the most unstable\cite{Drazin}, so it suffices to restrict this analysis to the 2D system. The inviscid assumption simplifies the calculation, although in physical systems at scales much smaller than those considered here, viscosity acts to remove energy from perturbations. The assumption of incompressibility is convenient because of the stabilizing effect of compressibility on shear flow instabilities\cite{Gerwin}. These assumptions allow the perturbed velocity to be written in terms of a stream function $\mathbf{v}_1 = \hat{y}\times\nabla \Phi(x,z) = (\partial \Phi / \partial z, 0, -\partial \Phi / \partial x)$. The perturbed vorticity is then entirely in the $-\hat{y}$ direction and is governed by the equation\cite{Drazin},
\begin{equation} \label{NLvorticity}
\frac{\partial}{\partial t}\nabla^2\Phi + U\frac{\partial}{\partial x}\nabla^2\Phi - \frac{\partial\Phi}{\partial x}\frac{d^2 U}{dz^2} + \frac{\partial\Phi}{\partial z}\frac{\partial}{\partial x}\nabla^2\Phi - \frac{\partial\Phi}{\partial x}\frac{\partial}{\partial z}\nabla^2\Phi = 0.
\end{equation}
This equation follows either from vorticity evolution in hydrodynamics or in MHD when the mean field is perpendicular to the flow and the fluctuations are electrostatic. We drop terms nonlinear in $\Phi$ and investigate normal modes of the form $\Phi(x,z,t) = \phi(k,z)\exp[ikx + i\omega(k)t]$, where $k = k^*d$ and $\omega = \omega^* d/U_0$. If we find that $\mathrm{Im}(\omega(k)) < 0$ for some mode at wavenumber $k$, then the mode is unstable and grows exponentially in time. If $\mathrm{Im}(\omega(k)) > 0$, the mode is stable and decays exponentially. If $\mathrm{Im}(\omega(k)) = 0$, the mode is marginally stable. We take Fourier modes in $x$ because Eq.~\eqref{NLvorticity} is homogeneous in $x$, but the dependence of $U$ on $z$ implies that Fourier modes in $z$ are not solutions to the linear equation. This significantly complicates the analysis of stable mode interactions, as discussed in the following section.

The linearized equation for the normal modes is\cite{DrazinHoward}
\begin{equation} \label{Lvorticity}
(\omega+kU)\left(\frac{d^2}{dz^2} - k^2\right)\phi - k\phi \frac{d^2U}{dz^2} = 0.
\end{equation}
Solutions of this system are well known\cite{Chandra}, but usually only the growth rate of the unstable mode and its eigenfunction are considered. We reexamine the problem to keep track of both the unstable and stable modes, in order to investigate their interaction through the nonlinearities in Eq.~\eqref{NLvorticity}.

Note that for the shear layer, $d^2U/dz^2$ is singular at $z = \pm 1$. For $|z| \neq 1$ however, $d^2U/dz^2 = 0$, so
\begin{displaymath}
(\omega+kU)\left(\frac{d^2}{dz^2} - k^2\right)\phi = 0
\end{displaymath}
(for $|z| \neq 1$). Solutions are given by either $\omega+kU = 0$ or $(d^2/dz^2 - k^2)\phi = 0$. While modes that satisfy the former are solutions of the system, we are interested in stable and unstable modes, which require $\mathrm{Im}(\omega) \neq 0$. Therefore we construct eigenmodes from $(d^2/dz^2 - k^2)\phi = 0$. It has been shown that for shear flow instabilities, the initial value calculation admits additional modes that decay algebraically\cite{Case}. While these modes potentially play a role in saturation of the instability and should be considered eventually, it makes sense to focus first on the interaction between the exponentially growing and decaying modes. Both the exponentially and algebraically decaying modes are ignored in quasilinear models of turbulence, so to show that these models overlook important, driving-scale features of the system it suffices to demonstrate the importance of stable modes.

Focusing on solutions of $(d^2/dz^2 - k^2)\phi = 0$, modes are given by
\begin{equation} \label{unsolved phi}
\phi(z) =
\begin{cases}
ae^{-|k|z} & z > 1 \\
e^{|k|z}+be^{-|k|z} & -1 < z < 1 \\
ce^{|k|z} & z < -1,
\end{cases}
\end{equation}
with the constants $a, b,$ and $c$ to be determined.

The flow profile $U(z)$ is continuous at the boundaries of the shear layer, which we assume to be fixed at $z = \pm 1$. Therefore $\phi$ must be continuous\cite{Chandra}, so $a$ and $c$ can each be written in terms of $b$. Although $U(z)$ and $\phi$ are continuous at $z = \pm 1$, the discontinuities in $dU/dz$ lead to discontinuities in $d\phi/dz$. The jump conditions that determine these discontinuities are obtained by integrating Eq.~\eqref{Lvorticity} from $-1-\epsilon$ to $-1+\epsilon$ and from $1-\epsilon$ to $1+\epsilon$, then taking $\epsilon \to 0$:
\begin{equation} \label{linearjump}
\lim_{\epsilon \to 0} (\omega \pm k)\frac{d\phi}{dz}\Big|_{\pm 1-\epsilon}^{\pm 1+\epsilon} \pm k\phi(\pm 1) = 0.
\end{equation}
After inserting Eq.~\eqref{unsolved phi}, these form two constraints on $b$ in terms of $\omega(k)$, which can be solved to obtain the dispersion relation,
\begin{equation} \label{dispersion}
\omega = \pm \frac{e^{-2|k|}}{2}\sqrt{e^{4|k|}(1-2|k|)^2-1}.
\end{equation}
\begin{figure}
\includegraphics[width=16cm]{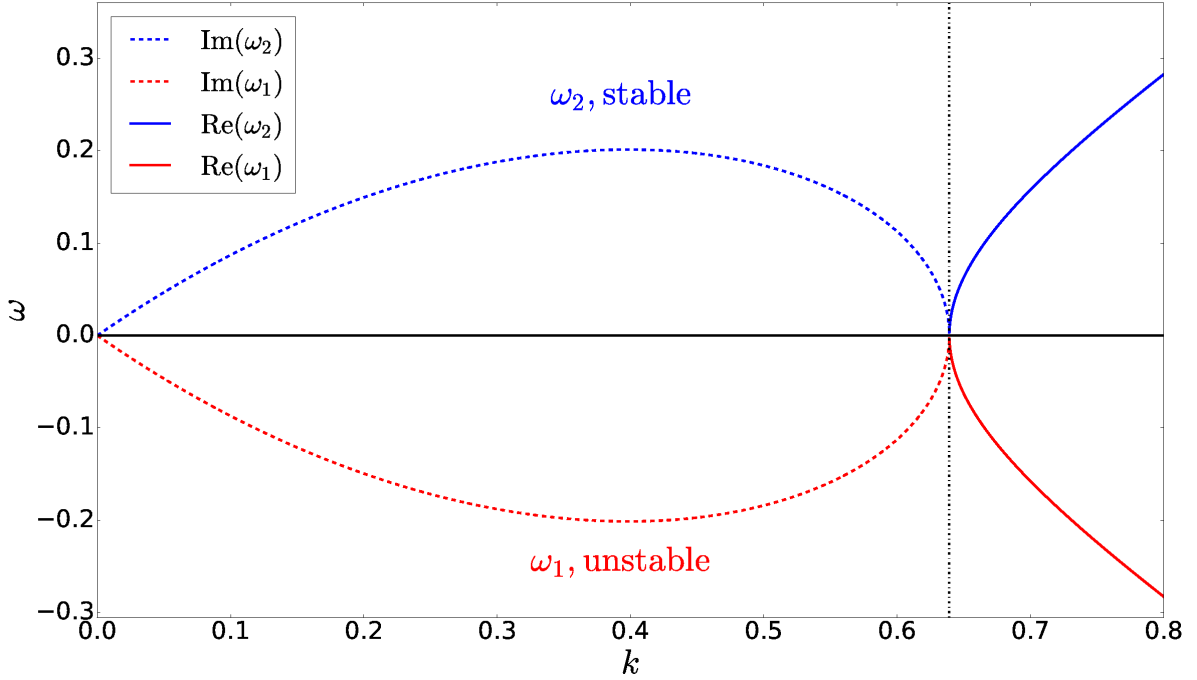}
\caption{Growth rate $\mathrm{Im}(\omega)$ and frequency $\mathrm{Re}(\omega)$ of the two modes for the inviscid shear layer. For $|k| \lesssim 0.64$ one mode is unstable and the other is stable, while for $|k| \gtrsim 0.64$ both modes are marginally stable.}\label{dispersion_plot}
\end{figure}%
Figure \ref{dispersion_plot} shows how the growth rates and frequencies depend on wavenumber. Note that $\omega^2 < 0$ for $0 < |k| < k_c$, where $k_c \approx 0.64$. For $k > k_c$, we shall refer to the negative and positive branches of $\omega$ as $\omega_1$ and $\omega_2$ respectively, noting that the reality condition requires $\omega_j(-k) = \omega_j^*(k)$. For $|k| < k_c$, we choose $\omega_1$ to be the unstable root and $\omega_2$ the stable one. Because $b$ depends on $\omega$ through Eq.~\eqref{linearjump} and the eigenmode structure $\phi(z)$ depends on $b$ through Eq.~\eqref{unsolved phi}, the two solutions $\omega_j$ correspond to two different eigenmodes $\phi_j(z)$. We identify $b_j$ and $\phi_j$ as the $b$ and $\phi$ corresponding to $\omega_j$. The eigenmodes are then
\begin{equation} \label{solved phi}
\phi_j(k,z) =
\begin{cases}
\left(e^{2|k|}+b_j\right)e^{-|k|z} & z > 1 \\
e^{|k|z}+b_je^{-|k|z} & -1 < z < 1 \\
\left(1+b_je^{2|k|}\right)e^{|k|z} & z < -1,
\end{cases}
\end{equation}
where  
\begin{equation}\label{bj}
b_j = e^{2|k|}\frac{2|k|(\omega_j+k)-k}{k}
\end{equation}
satisfies $b_1(k) = b_2(-k) = b_2^*(k)$ for $|k|<k_c$, and $b_j(k) = b_j(-k) = b_j^*(k)$ for $|k|>k_c$.
\begin{figure}
	\includegraphics[width=16cm]{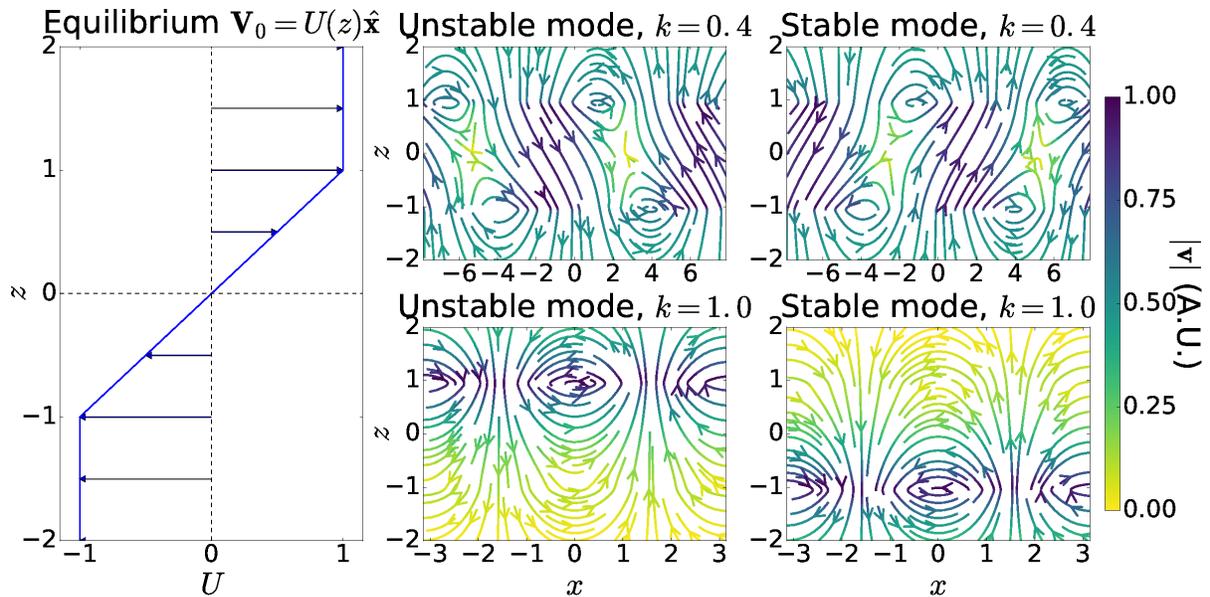}
	\caption{Equilibrium (left column) compared with velocity profiles of the unstable $\phi_1$ (middle column) and the stable $\phi_2$ (right column) at wavenumbers $k = 0.4$ (top row) and $k=1$ (bottom row) plotted over one wavelength in $x$ and from $z=-2$ to $z=2$. Streamlines are plotted with color indicating flow speed. The first row is in the unstable range, where $\phi_1$ grows exponentially while $\phi_2$ decays exponentially. The second row is a marginally stable wavenumber, where both $\phi_1$ and $\phi_2$ oscillate without any growth [see Fig.~\ref{dispersion_plot}].}
	\label{eigenmode_plot}
\end{figure}
For $\omega^2<0$, the eigenmodes are nearly identical but satisfy $\phi_1(k,z) = \phi^*_2(k,z)$. Figure \ref{eigenmode_plot} shows the flows corresponding to these eigenmodes for four wavenumbers sampling the unstable and stable ranges. Previous work has shown that the physical mechanisms for instability of $\phi_1$ and stability of $\phi_2$ can be understood in terms of resonant vorticity waves in both the hydrodynamic\cite{Baines} and MHD\cite{Heifetz} systems.

In standard descriptions of turbulence and quasilinear transport calculations, it is common practice to neglect stable modes given their exponential decay from a small initial value. In this paper we account for the nonlinear drive of the stable mode by the unstable mode and investigate its impact on the evolution of the system.

\section{Eigenmode Projection}
In previous calculations of stable mode excitation\cite{Terry2006,Makwana}, fluctuations from equilibrium were represented by a vector $\mathbf{f}(\mathbf{k},t)$ whose components were the Fourier-transformed dynamical quantities. Because the systems were homogeneous, the linearized, Fourier-transformed PDEs became ODEs of the form $\dot{\mathbf{f}} = \mathcal{D}\mathbf{f}$ with the dynamics at each wavenumber $\mathbf{k}$ linearly decoupled. Thus, the eigenmodes of the system were the eigenvectors $\mathbf{f}_j$ of the $N\times N$ matrix $\mathcal{D}$, and arbitrary states could be expanded as linear combinations of the eigenvectors:
\begin{equation} \label{2006 expansion}
\mathbf{f}(\mathbf{k},t) = \sum_{j=1}^{N}\beta_j(\mathbf{k},t)\mathbf{f}_j(\mathbf{k},t),
\end{equation}
where $\beta_j(\mathbf{k},t)$ is the component of $\mathbf{f}$ in the eigenmode basis. Also called eigenmode amplitudes, the functions $\beta_j$ are not specified by the solutions of the linearized equations except through an initial condition. Under linear evolution the stable modes subsequently decrease to insignificance. However, the full nonlinear ODEs can readily be written in terms of the eigenmodes by substituting the eigenmodes for the dynamical quantities using Eq.~\eqref{2006 expansion}. From there, separate equations for each $\dot{\beta}_j$ can be derived. These equations for $\dot{\beta}_j$ are equivalent to the original PDEs, but they describe the nonlinear evolution of the system in terms of the eigenmode amplitudes. We refer to this process, both the expansion of the perturbations and the manipulation of their governing equations, as an eigenmode decomposition. The equations for $\dot{\beta}_j$ provide powerful insight into the system. The nonlinearities that couple the dynamical fields at different scales become nonlinearities that couple eigenmodes at different wavelengths. Thus, it was shown (and borne out by many simulations\cite{Terry2006}) that despite decaying in the linear regime, the stable modes are nonlinearly driven by the unstable modes.

In these previous calculations, the homogeneous nature of the system made the set of linear eigenmodes a complete basis: at every time $t$ and wavevector $\mathbf{k}$, the state vector $\mathbf{f}$ could be expanded in a basis of the eigenmodes [i.e.~Eq.~\eqref{2006 expansion}]. Due to the inhomogeneity of the present system, the linear solutions are not simply Fourier modes in $z$, so this system does not readily lend itself to the vector representation of Ref.~\cite{Terry2006}. Moreover, Eq.~\eqref{NLvorticity} admits only two eigenmodes which are expected not to span arbitrary perturbations that satisfy the boundary conditions\cite{Case}. So the true state of the system cannot be written exactly in the form of Eq.~\eqref{2006 expansion} with $N = 2$. In order to properly describe the evolution of the system given an arbitrary initial condition, the system could be expanded in appropriate orthogonal polynomials or investigated as an initial value problem with additional time-dependent basis functions that are linear solutions of the problem. Previous work has demonstrated that for inviscid shear flows, the initial value calculation leads to the ``discrete" eigenmodes with time-dependence $\exp[i\omega t]$ described in the previous section, and an additional set of ``continuum" modes\cite{Case}. These continuum modes either oscillate with frequency $k$ or decay algebraically. For the present calculation we only consider perturbations that can be expressed as linear combinations of the two discrete eigenmodes $\phi_1$ and $\phi_2$, representing a truncation of the complete system. If we are able to demonstrate a significant impact from $\phi_2$, that suffices to demonstrate the importance of stable modes, relative to existing models that only consider the unstable mode.

By focusing on perturbations that are linear combinations of $\phi_1$ and $\phi_2$ (i.e.~limiting ourselves to the subspace spanned by $\phi_1$ and $\phi_2$), the vector representation and invertible linear transformation between the state of the system and the eigenmode amplitudes of Ref.~\cite{Terry2006} can be recovered. Consequently, the governing Eq.~\eqref{NLvorticity} can be manipulated to derive nonlinear equations that describe the evolution of the eigenmode amplitudes and their interactions. The method relies on the jump conditions given in Eq.~\eqref{linearjump}. Since the jump conditions for one eigenmode differ from those for the other eigenmode, one can form an invertible map between the discontinuity of $d\phi/dz$ at each interface and the amplitude of each eigenmode. Additionally, because there are two jump conditions that will serve as our dynamical quantities, only the two eigenmodes of the previous section are needed to construct an invertible map between eigenmodes and dynamical quantities. To derive equations describing the nonlinear interaction between the eigenmodes, we start by deriving nonlinear jump conditions.

First, let $\hat{\phi}(k,z,t) = \mathcal{F}[\Phi(x,z,t)]$ be the Fourier transformed stream function, and assume
\begin{equation}\label{phihat combo}
\hat{\phi}(k,z,t) = \beta_1(k,t)\phi_1(k,z) + \beta_2(k,t)\phi_2(k,z).
\end{equation}
The nonlinear jump conditions are obtained by performing the same steps that led to Eq.~\eqref{linearjump} without dropping nonlinear terms (and explicitly taking the Fourier transform rather than assuming normal modes). Taking the Fourier transform and integrating from $\pm 1 - \epsilon$ to $\pm 1 + \epsilon$ with $\epsilon \to 0$ yields
\begin{equation}\label{deltahat}
\frac{\partial}{\partial t}\hat{\Delta}_{\pm} \pm ik\hat{\Delta}_{\pm} \pm ik\hat{\phi}(k,\pm 1) + \lim\limits_{\epsilon\to 0}ik\int \limits_{-\infty}^{\infty}\frac{dk'}{2\pi}\left[\frac{d}{dz}\hat{\phi}(k',z)\frac{d}{dz}\hat{\phi}(k'',z)\right]_{\pm 1-\epsilon}^{\pm 1+\epsilon} = 0,
\end{equation}
where $k'' \equiv k-k'$, while
\begin{align*}
\hat{\Delta}_{\pm}(k,t) &\equiv \lim\limits_{\epsilon\to 0}\left[ \frac{d}{dz}\hat{\phi}(k,\pm 1 + \epsilon,t) - \frac{d}{dz}\hat{\phi}(k,\pm 1 - \epsilon,t)\right]\\
&= \beta_1(k,t)\Delta_{\pm 1}(k) + \beta_2(k,t)\Delta_{\pm 2}(k)
\end{align*}
and
\begin{equation*}
\Delta_{\pm j}(k) \equiv \lim\limits_{\epsilon\to 0}\left[ \frac{d}{dz}\phi_j(k,\pm 1 + \epsilon) - \frac{d}{dz}\phi_j(k,\pm 1 - \epsilon)\right]
\end{equation*}
are the discontinuities in $d\hat{\phi}/dz$ and $d\phi_j/dz$ at $z = \pm 1$. With $\hat{\phi}$ given by Eq.~\eqref{phihat combo} and $\phi_j$ given by Eq.~\eqref{solved phi}, one can show that
\begin{equation*}
\phi_j(k,1) = \frac{-\Delta_{+ j}}{2|k|} - \frac{\Delta_{- j}}{2|k|e^{2|k|}},
\end{equation*}
and
\begin{equation*}
\phi_j(k,-1) = \frac{-\Delta_{+ j}}{2|k|e^{2|k|}} - \frac{\Delta_{- j}}{2|k|}.
\end{equation*}
The $\hat{\phi}(k,\pm 1)$ term in Eq.~\eqref{deltahat} can then be written in terms of $\hat{\Delta}_{\pm}$ to yield
\begin{equation}\label{deltahat corrected}
\frac{\partial}{\partial t}\begin{pmatrix}\hat{\Delta}_+\\\hat{\Delta}_-\end{pmatrix} = \mathcal{D}\begin{pmatrix}\hat{\Delta}_+\\\hat{\Delta}_-\end{pmatrix} + \begin{pmatrix}N_+\\N_-\end{pmatrix},
\end{equation}
with
\begin{equation}\label{Dmatrix}
\mathcal{D} = ik\begin{pmatrix}
\frac{1}{2|k|} - 1 & \frac{e^{-2|k|}}{2|k|}\\
\frac{-e^{-2|k|}}{2|k|} & -\frac{1}{2|k|} + 1\\
\end{pmatrix},
\end{equation}
and $N_{\pm}$ representing the nonlinearities in Eq.~\eqref{deltahat}. Note that taking $N_{\pm} \to 0$ and $\partial / \partial t \to i\omega$ reduces this to the linear system solved in the previous section.

We now have all of the necessary tools to treat this system in a manner similar to the previously-mentioned calculations\cite{Terry2006,Makwana}. Using our definitions for $\hat{\Delta}_{\pm}$ and $\Delta_{\pm j}$, the $z$-derivative of Eq.~\eqref{phihat combo} evaluated between $z = \pm 1 + \epsilon$ and $z = \pm 1 - \epsilon$ with $\epsilon \to 0$ is
\begin{equation}\label{Mdef}
\begin{pmatrix}\hat{\Delta}_+\\\hat{\Delta}_-\end{pmatrix} = \mathbf{M}\begin{pmatrix}\beta_1\\\beta_2\end{pmatrix},
\end{equation}
where
\begin{equation}\label{Mmatrix}
\mathbf{M} = \begin{pmatrix}
\Delta_{+ 1} & \Delta_{+ 2}\\
\Delta_{- 1} & \Delta_{- 2}\\
\end{pmatrix} = -2|k|e^{|k|}\begin{pmatrix}
1 & 1\\
b_1 & b_2\\
\end{pmatrix},
\end{equation}
and $b_j$ is given in Eqn.~\eqref{bj}. Equation \eqref{Mdef} is equivalent to Eq.~\eqref{2006 expansion}: for this calculation, the dynamical quantities that we use to specify the state of the system are $\hat{\Delta}_{\pm}$, and their eigenmode structure is given by the columns of the matrix $\mathbf{M}$. The governing nonlinear PDE, Eq.~\eqref{NLvorticity} has been rewritten as a system of nonlinear ODEs, Eq.~\eqref{deltahat corrected}. The linearized system of ODEs (Eq.~\eqref{deltahat corrected} with $N_{\pm} \to 0$) can be diagonalized: substituting $\hat{\Delta}_{\pm}$ for $\beta_j$ via Eq.~\eqref{Mdef} and multiplying by $\mathbf{M}^{-1}$ on the left gives
\begin{equation}\label{Leigenmode}
\begin{pmatrix}\dot{\beta}_1\\\dot{\beta}_2\end{pmatrix} = \mathbf{M}^{-1}\mathcal{D}\mathbf{M}\begin{pmatrix}\beta_1\\\beta_2\end{pmatrix},
\end{equation}
where the matrix $\mathbf{M}^{-1}\mathcal{D}\mathbf{M}$ is diagonal with entries $i\omega_j$.

The nonlinear interactions between the eigenmodes can now be investigated. Applying the steps that led to Eq.~\eqref{Leigenmode} to the full, nonlinear Eq.~\eqref{deltahat corrected} yields
\begin{equation}\label{NLeigenmode}
\begin{pmatrix}\dot{\beta}_1\\\dot{\beta}_2\end{pmatrix} = \mathbf{M}^{-1}\mathcal{D}\mathbf{M}\begin{pmatrix}\beta_1\\\beta_2\end{pmatrix} + \mathbf{M}^{-1}\begin{pmatrix}N_+\\N_-\end{pmatrix},
\end{equation}
where, again, $N_{\pm}$ are the nonlinearities in Eq.~\eqref{deltahat}. Using Eq.~\eqref{phihat combo} and the forms for $\phi_j$ given by Eq.~\eqref{solved phi}, $N_{\pm}$ can be written in terms of products of the form $\beta_i \beta_j$ with $i,j$ each taking values $1,2$. Equation \eqref{NLeigenmode} then becomes
\begin{equation}\label{unstablemode}
\begin{split}
\dot{\beta}_1(k) = i\omega_1(k)\beta_1(k) + \int \limits_{-\infty}^{\infty}\frac{dk'}{2\pi}\bigg[ &C_1(k,k') \beta_1(k')\beta_1(k'') + C_2(k,k') \beta_1(k')\beta_2(k'')\\
+ &C_3(k,k') \beta_1(k'')\beta_2(k') + C_4(k,k')\beta_2(k')\beta_2(k'')\bigg],
\end{split}
\end{equation}
\begin{equation}\label{stablemode}
\begin{split}
\dot{\beta}_2(k) = i\omega_2(k)\beta_2(k) + \int \limits_{-\infty}^{\infty}\frac{dk'}{2\pi}\bigg[ &D_1(k,k')\beta_1(k')\beta_1(k'') + D_2(k,k')\beta_1(k')\beta_2(k'')\\
+ &D_3(k,k')\beta_1(k'')\beta_2(k') + D_4(k,k')\beta_2(k')\beta_2(k'')\bigg].
\end{split}
\end{equation}
The coefficients $C_j,D_j$ arise from writing the nonlinearities $N_{\pm}$ in the basis of the linear eigenmodes, so their functional forms include information about both the linear properties of the system and the nonlinearities $N_{\pm}$. The exact expressions for $C_j, D_j$ are given in the Appendix, where it is shown that $C_2(k,k') = C_3(k,k-k')$, so that the $C_3$ integral is equal to the $C_2$ integral. Equations \eqref{unstablemode} and \eqref{stablemode} are equivalent to Eq.~\eqref{deltahat corrected}, but they directly show how $\beta_1$ and $\beta_2$ interact.

\begin{figure}
	\includegraphics[width=16cm]{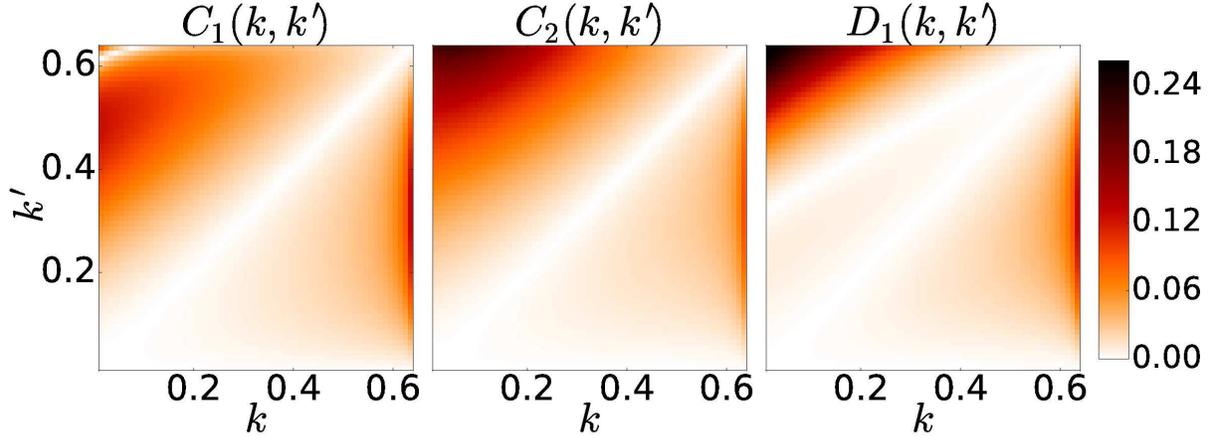}
	\caption{Three of the eight nonlinear coupling coefficients in Eqs.~\eqref{unstablemode} and \eqref{stablemode}, $C_1, C_2$, and $D_1$, evaluated over the most relevant scales. Color indicates absolute values of the coefficients. The coefficients are all roughly the same magnitude, indicating significant coupling between stable and unstable eigenmodes.}
	\label{CDplot}
\end{figure}

An analogy can be made here to the use of Els{\"a}sser variables in incompressible, homogeneous MHD turbulence, which are a familiar example of an eigenmode decomposition that makes explicit the nonlinear interaction of the linear eigenmodes. The linearized equations have as solutions counterpropagating, noninteracting waves of the form $\mathbf{z}^{\pm} = \mathbf{v} \pm \mathbf{b}/(4\pi\rho_0)^{1/2}$. Expressing the nonlinear equations in terms of $\mathbf{z}^{\pm}$, the nonlinearity in the equation for $\partial \mathbf{z}^{\pm}/\partial t$ is $\mathbf{z}^{\mp}\cdot\nabla\mathbf{z}^{\pm}$, which describes the nonlinear interactions between linearly noninteracting modes. In the present calculation, the linearly noninteracting $\phi_1, \phi_2$ are comparable to $\mathbf{z}^{\pm}$, and the terms proportional to $\beta_1(k')\beta_2(k'')$ and $\beta_1(k'')\beta_2(k')$ are comparable to $\mathbf{z}^{\mp}\cdot\nabla\mathbf{z}^{\pm}$. However, unlike the $\mathbf{z}^{\pm}$ equations, the $\dot{\beta}_j$ equations include other nonlinear terms that are proportional to $\beta_1(k')\beta_1(k'')$ and $\beta_2(k')\beta_2(k'')$. If all of the nonlinearities are zero except for the $C_1$ term, then the evolution of $\beta_1(k)$ is just a combination of its linear drive $i\omega_1(k)$ and three-wave interactions with $\beta_1(k')$ and $\beta_1(k-k')$, allowing $\phi_1$ to saturate through a Kolmogorov-like cascade to smaller scales. This is effectively the assumption of standard quasilinear calculations of momentum transport, where only $\phi_1,\omega_1$ are considered. Figure \ref{CDplot} shows some of the nonlinear coupling coefficients plotted over a range of wavenumbers. Since $D_1, C_2$, and $C_3$ are not identically zero, there is some interaction between the eigenmodes. Systems where such interactions have been identified are all gyroradius-scale, quasihomogeneous systems driven by drift-wave instabilities\cite{Terry2006,Makwana}. Equations \eqref{unstablemode} and \eqref{stablemode} represent a demonstration that these interactions occur for larger-scale, inhomogeneous plasmas and neutral fluids.
\section{The Threshold Parameter}
By comparing the nonlinearities that transfer energy to stable modes with those that cause the Kolmogorov-like cascade of energy to small scales, one can investigate how important stable modes are in instability saturation. A quantity known as the threshold parameter $P_t$ has been used to evaluate the relative importance of the stable eigenmodes in situations where instability saturation is described by eigenmode-projected ODEs. The threshold parameter $P_t$ estimates the relative importance in saturation of the nonlinearities responsible for energy transfer to the stable mode versus the nonlinearity of the forward cascade\cite{Terry2006}. If $P_t$ is small compared to unity, it indicates that the instability saturates via a Kolmogorov-like transfer of energy to smaller scales, and only the term in Eq.~\eqref{2006 expansion} corresponding to the most unstable eigenmode needs to be included to accurately describe the system. If $P_t \gtrsim 0.3$, it was found that the transfer of energy from the unstable mode to other modes at similar scales is an important mechanism in saturation. In that case, additional terms in Eq.~\eqref{2006 expansion} must therefore be included\cite{Makwana}.

The quantity $P_t$ is the ratio of the $C_1\beta_1\beta_2$ and $C_2\beta_1\beta_1$ terms in Eq.~\eqref{unstablemode} and therefore includes information about both linear and nonlinear properties of the system. In previous work\cite{Terry2006,Makwana}, simplifying assumptions -- such as treating growth rates $\gamma_j = -\mathrm{Im}(\omega_j)$ as independent of wavenumber -- allowed the threshold parameter to be written as
\begin{equation}\label{Ptold}
P_t = \frac{2 D_1 C_2}{C_1^2(2-\gamma_2/\gamma_1)}
\end{equation}
for $\gamma_2<0$. This form of $P_t$ is useful because it illustrates how $P_t$ depends on different parameters of the system: the size of $P_t$ relative to unity is determined by the ratios $D_1C_2/C_1^2$ and $\gamma_1/\gamma_2$. When the former is small, stable modes are only weakly coupled to unstable modes and have little impact on saturation dynamics. When the latter is small, stable modes decay too quickly to achieve significant amplitude by the time the instability saturates unless $D_1C_2/C_1^2 \gg 1$ and compensates. Previous work evaluated this form of $P_t$ in several systems and found that whenever $P_t$ is at least a few tenths, energy transfer to stable modes becomes comparable to the energy injection rate of the instability\cite{Makwana}. Note that in the system considered here $|\gamma_1/\gamma_2| = 1$, and numerically evaluating $C_j,D_j$ shows that $D_1$ and $C_2$ are of the same order as or even larger than $C_1$ [see Fig.~\ref{CDplot}]. These features alone yield $P_t \approx 0.7$, which implies stable modes are important for KH saturation.

Here we extend previous analyses of $P_t$ by including the full wavenumber dependence of $\gamma_j, C_j$, and $D_j$. Consider the evolution of the system from a small initial amplitude $\beta_i$. When amplitudes are small every nonlinear term is negligible, so the dynamics are linear with $\beta_2$ decaying and $\beta_1$ growing exponentially at every wavenumber. Eventually couplings in $\int (dk'/2\pi) D_1(k,k')\beta_1(k')\beta_1(k-k')$ dominate in Eq.~\eqref{stablemode}. This occurs in the linear phase, before saturation, because nonlinearities dominate the decaying linear response of $\beta_2$ long before matching the growing linear response of $\beta_1$. Thus, Eq.~\eqref{stablemode} can be approximated as
\begin{equation}\label{parametricstable}
\dot{\beta}_2(k) = i\omega_2(k)\beta_2(k) + \int \limits_{-\infty}^{\infty}\frac{dk'}{2\pi}D_1(k,k')\beta_1(k')\beta_1(k'').
\end{equation}
Note that for these times $\beta_2 \ll \beta_1$ therefore the $D_1$ terms are the largest of the $D_j$ terms.
Since the $C_j$ nonlinearities have not reached the amplitudes of the growing linear terms, $\beta_1$ can be approximated as $\beta_i \exp[i \omega_1 t]$. These approximations are referred to as the parametric instability approximations\cite{Terry2006}. Then Eq.~\eqref{parametricstable} is solved by
\begin{equation}\label{parametricsolution}
\beta_2(k,t) = \int \limits_{-\infty}^{\infty}\frac{dk'}{2\pi} \frac{D_1(k,k')\beta_i^2}{i\left(-\omega_2(k)+\omega_1(k')+\omega_1(k-k')\right)} \left[ e^{i(\omega_1(k')+\omega_1(k-k'))t} - e^{i\omega_2(k)t} \right] + \beta_i e^{i \omega_2(k) t}.
\end{equation}
In assessing $P_t$ the above integral is only taken over unstable wavenumbers, as they drive $\beta_2$ more strongly than marginally stable modes.

\begin{figure}
	\includegraphics[width=16cm]{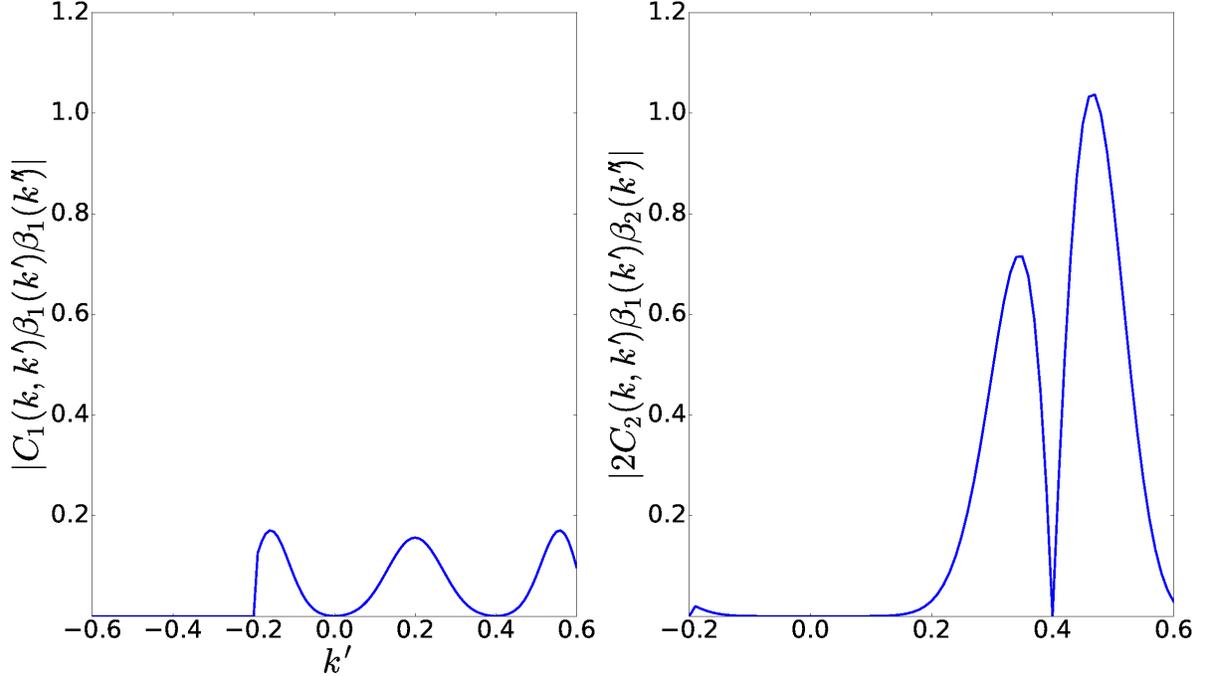}
	\caption{Nonlinear terms in Eq.~\eqref{unstablemode} at saturation for $k=0.4$ and $\beta_i=0.01,$ with $k'$ and $k-k'$ ranging from $-0.6$ to $0.6$. The $C_1$ term is responsible for the Kolmogorov-like saturation of the instability by energy transfer to unstable modes at smaller wavelengths. The $C_2$ term represents the previously-neglected coupling between unstable modes at $k$ and $k'$ with stable modes at $k''$. The threshold parameter is evaluated by dividing the peak value of the $C_2$ term by the peak value of the $C_1$ term. Here we find $P_t \approx 6$, indicating that stable modes are important in KH saturation.}
	\label{PTscan}
\end{figure}
To evaluate $P_t$, the ratio of the largest $\beta_1\beta_2$ term and the largest $\beta_1\beta_1$ term in Eq.~\eqref{unstablemode} is taken at the time of saturation $t_s$:
\begin{equation}\label{Ptnew}
P_t = \left[\frac{\max|2C_2\beta_1(k')\beta_2(k'')|}{\max|C_1\beta_1(k')\beta_1(k'')|}\right]_{t=t_s},
\end{equation}
where $t_s$ is defined as the time at which one of the nonlinearities in Eq.~\eqref{unstablemode} reaches the same amplitude as the linear term. Figure \ref{PTscan} shows the size of these terms at saturation for $k=0.4$ with an initial amplitude of $\beta_i = 0.01$. We choose $k=0.4$ because it is the most unstable wavenumber and is therefore the wavenumber of the most dominant unstable mode leading into saturation. From Fig.~\ref{PTscan}, it is inferred that $P_t \approx 6$, indicating that even before the nominal saturation time energy transfer to stable modes has become as important to the saturation of the unstable mode at $k=0.4$ as the Kolmogorov-like transfer to unstable modes at other scales.

In previous calculations of $P_t$, the parameter was independent of the initial amplitude $\beta_i$ (which is assumed to be the same for each $k$). However, in the above evaluation of $P_t$, we do find that it depends on $\beta_i$; for instance, reducing $\beta_i$ to $0.001$ yields $P_t \approx 15$. In previous calculations, the lack of dependence of $P_t$ on $\beta_i$ is an artifact of treating growth rates as independent of wavenumber\cite{Terry2006}. In considering Eq.~\eqref{unstablemode} for the most unstable wavenumber, both $\beta_1(k')$ and $\beta_1(k'')$ were assumed to grow at the same rate as the most unstable mode, when in fact three-wave interactions require $k \neq k'$. When including wavenumber dependence, these nonlinear terms will necessarily grow at less than twice the peak growth rate. On the other hand, stable modes near $k=0$ can be driven by $D_1\beta_1\beta_1$ terms where one of the driving modes is near $k=0.4$ and the other is near $k=-0.4$. Thus, our inclusion of the wavenumber dependence of $\omega_j$ causes $\beta_2$ to grow large enough that Eq.~\eqref{parametricstable} becomes invalid before saturation time. This makes the precise value of $P_t$ less meaningful, as the stable modes have grown so large that the approximations made in obtaining $P_t$ are invalid. However, the size of $\beta_2$ relative to $\beta_1$ and the comparable amplitudes of $C_2$ and $C_1$ imply $P_t \gtrsim 1$, and therefore $P_t \gtrsim 0.3$ is still well satisfied.

The above nonlinear analysis demonstrates that energy transfer to stable modes is significant relative to energy transfer to smaller scales, modifying the usual understanding of instability saturation by a cascade to small scales. The analysis employs approximations, hence it is instructive to consider a second, complementary form of approximate nonlinear analysis based on a three-wavenumber truncation of Eqs.~\eqref{unstablemode} and \eqref{stablemode}. Such a calculation complements the $P_t$ analysis because it makes different assumptions. The $P_t$ analysis makes parametric instability approximations when considering the evolution of $\beta_2$ (c.f.~Eq.~\eqref{parametricsolution}), but samples a broad continuum of wavenumbers. On the other hand, a three-wavenumber truncation makes no assumptions about the evolution of the modes, and instead limits the system to only three wavenumbers that are evolved according to Eqs.~\eqref{unstablemode} and \eqref{stablemode}.
\begin{figure}
	\includegraphics[width=16cm]{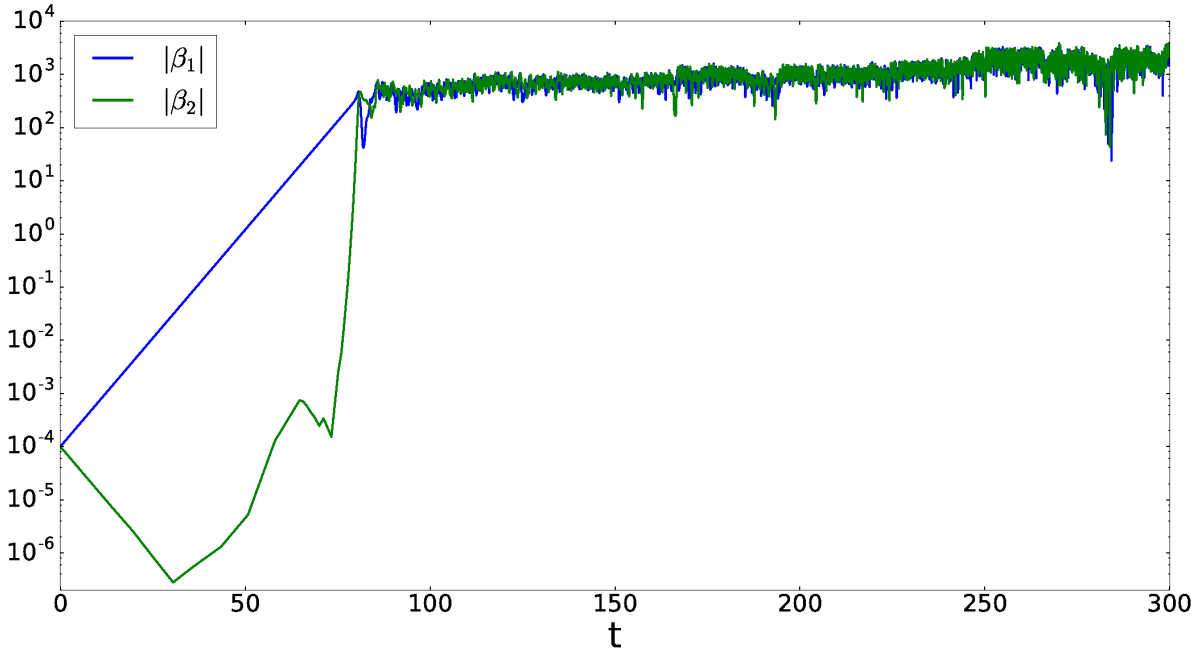}
	\caption{Time evolution of $|\beta_j(k,t)|$ for a three-wavenumber truncation with $k=0.3, k'=0.9$, and $k-k'=-0.6$. As expected from the $P_t$ analysis, the stable mode decays linearly, then is nonlinearly pumped to an amplitude that is comparable to the unstable mode.}
	\label{ThreeModePlot}
\end{figure}

The result of a three-wavenumber truncation is plotted in Fig.~\ref{ThreeModePlot}, showing the time evolution of $\beta_j(k,t)$ obtained by solving Eqs.~\eqref{unstablemode} and \eqref{stablemode} numerically with only interactions between $k=0.3$, $k'=0.9$, and $k-k'=-0.6$ considered. The linear growth phase of $\beta_1$ is clearly seen, as is the linear decay and nonlinear driving of $\beta_2$. The linear growth phase for $\beta_1$ ends with both eigenmodes reaching comparable amplitudes, consistent with the $P_t$ analysis. Once the stable mode reaches a value that is comparable to the unstable mode there is continuous exchange of energy between the two modes. The saturation levels slowly grow as $t \to \infty$. That can be understood as a consequence of the inviscid dynamics in a three-mode system, in that previous work has demonstrated that a necessary condition for bounded solutions to three-mode truncations is that the sum of the growth rates is negative\cite{Terry1982}. Without viscosity, the present system does not meet the necessary condition. Note that the time scale for nonlinear energy exchange is very short compared to the time scale of the saturation level increase, strongly suggesting that the nonlinearities of Eqs.~\eqref{unstablemode} and \eqref{stablemode} conserve energy. This calculation demonstrates that the system can saturate by energy transfer to stable modes, and shows that the assumptions made regarding the growth of $\beta_1$ and $\beta_2$ in the $P_t$ analysis are reasonable.

\begin{figure}
	\includegraphics[width=16cm]{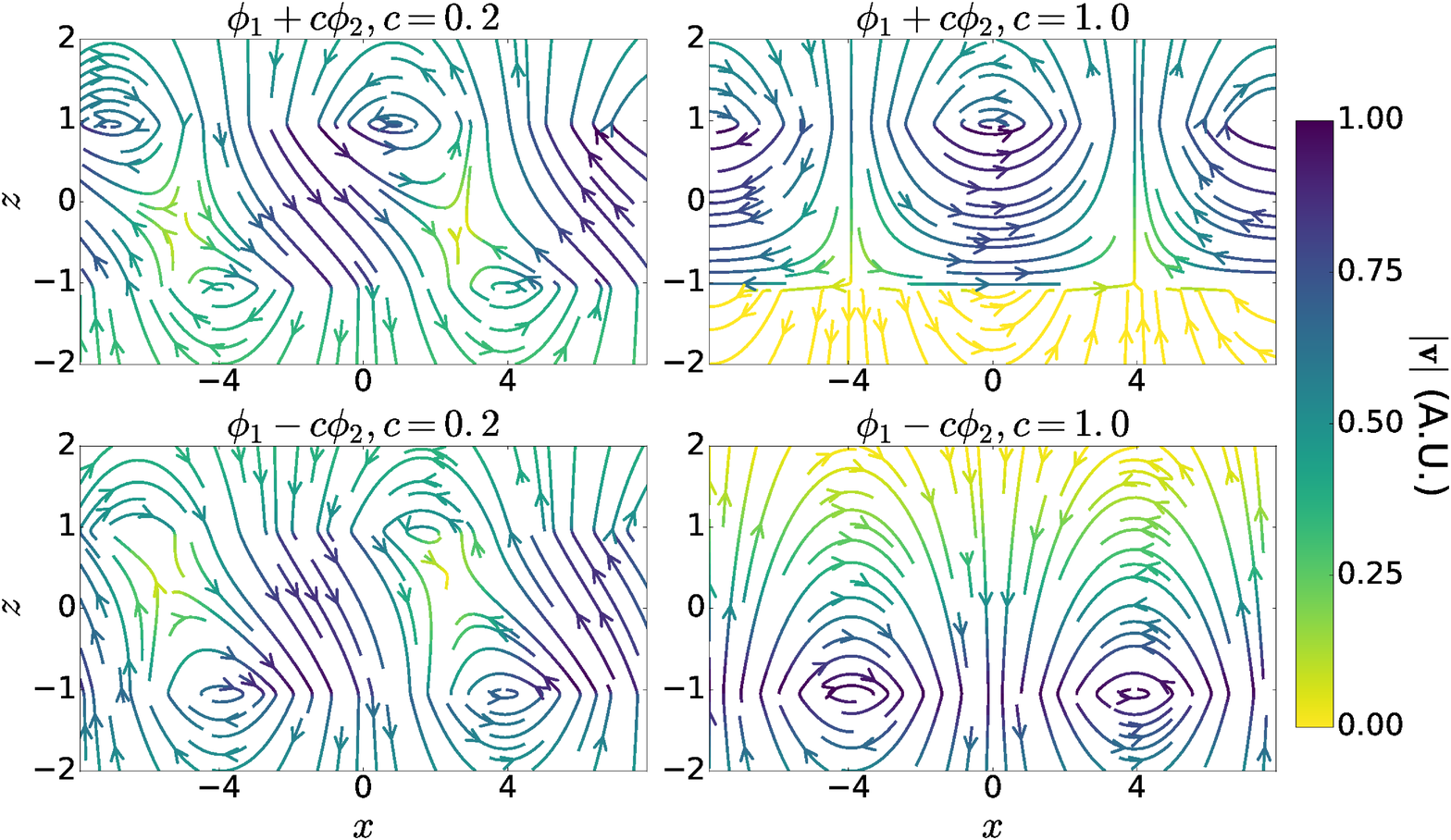}
	\caption{Examples of superpositions of stable and unstable modes at $k=0.4$ plotted over one wavelength in $x$ and from $z=-2$ to $z=2$ (cf.~Fig.~\ref{eigenmode_plot}). In the right column, the unstable and stable modes have an equal contribution to the overall flow. In the top and bottom rows, the relative phase between the two modes is $+\pi$ and $-\pi$, respectively.}
	\label{summed_mode_plot}
\end{figure}

As an illustration of the effect of finite $\beta_2$ on the fluctuating flow, Fig.~\ref{summed_mode_plot} shows the flows arising from linear combinations of $\beta_1$ and $\beta_2$ with the weight of $\beta_2$ varied. The flow arising purely from the unstable mode is strikingly different from the flow that combines $\beta_1$ and $\beta_2$ with equal weights. Regions of hyperbolic flow appear to be less likely for the equally weighted combination, suggesting that secondary structure generation and cascading may be weakened when the stable mode is excited. This will be the subject of future research.\\
\section{Momentum Transport}
Reynolds stresses and the associated momentum transport due to unstable modes tend to broaden the original flow profile. Here we show that stable modes have the potential to reduce the broadening of the profile. The transport of momentum in the $x$ direction across the interface at $z=1$ is found by integrating the $x$-component of the divergence of the stress tensor $\tau_{ij} = \langle v_{1i} v_{1j} \rangle$ across the interface. Integrating $d\tau_{xz}/dz$ across the interface gives
\begin{equation*}
S = -\lim\limits_{\epsilon \to 0} \int \limits_{1-\epsilon}^{1+\epsilon}dz\langle v_{1x}v_{1z} \rangle = -\lim\limits_{\epsilon \to 0} \int \limits_{1-\epsilon}^{1+\epsilon}dz\frac{d}{dz}\langle \frac{d\Phi}{dz}\frac{\partial \Phi}{\partial x}\rangle
\end{equation*}
where $\langle  \rangle$ denotes averaging in $x$, while $\mathbf{v}_1$ is the perturbed velocity. Taking $\Phi = \mathcal{F}^{-1}[\hat{\phi}]$ with $\hat{\phi} = \beta_1\phi_1 + \beta_2\phi_2$ gives
\begin{equation}\label{shear}
\begin{split}
S = \int \limits_{-\infty}^{\infty}\frac{dk}{2\pi}4k^2e^{2|k|} \bigg[  &\mathrm{Im}(\omega_1^*)|\beta_1|^2 + \mathrm{Im}(\omega_2^*)|\beta_2|^2\\
+ &\mathrm{Im}\left[(\omega_2^*+k)\beta_1\beta_2^*\right] + \mathrm{Im}\left[(\omega_1^*+k)\beta_2\beta_1^*\right]\bigg].
\end{split}
\end{equation}
When the stable modes are ignored, only the first term contributes to $S$. The coefficient $4k^2e^{2|k|}$ is positive, and Eq.~\eqref{dispersion} shows that $\mathrm{Im}(\omega^*_1) \leq 0$ and $\mathrm{Im}(\omega^*_2) = -\mathrm{Im}(\omega^*_1)$, indicating that the transport due to unstable modes alone is negative, and the second term acts against the first to reduce $|S|$. Clearly the amplitude of $\beta_2(k)$ relative to $\beta_1(k)$ has a significant impact on the momentum transport in this system. The relative phase between $\beta_2(k)$ and $\beta_1(k)$ determines the contribution of the last two terms. If $|\beta_2(k)| = |\beta_1(k)|$, then the first two terms cancel and the transport is entirely determined by the last two terms. Analysis of other systems shows there are situations where eigenmode cross correlations significantly affect transport\cite{Baver,Terry2009}.

To determine the actual properties of $S$, it is necessary to solve Eqs.~\eqref{unstablemode} and \eqref{stablemode} for $\beta_j(k)$ and integrate Eq.~\eqref{shear}, either analytically or numerically. This is beyond the scope of the present paper, but will be considered in the future. In lieu of such solutions, we construct an estimate of the ratio $|\beta_2(k)|/|\beta_1(k)|$ using the threshold parameter.

In the previous section the threshold parameter was defined as the ratio of the maximum amplitudes of the $C_2$ terms and the $C_1$ terms in Eq.~\eqref{unstablemode} at the onset of saturation. An estimate of $|\beta_2(k)|/|\beta_1(k)|$ in terms of $P_t$ is obtained by taking
\begin{equation*}
P_t \sim \frac{|2C_2\beta_1(k')\beta_2(k'')|}{|C_1\beta_1(k')\beta_1(k'')|} = 2\left|\frac{C_2}{C_1}\right|\frac{|\beta_2(k'')|}{|\beta_1(k'')|} \sim 2\left|\frac{\beta_2(k)}{\beta_1(k)}\right|.
\end{equation*}
While the threshold parameter estimates the relative amplitudes of the modes, it does not capture information about their cross-phase. Taking $\beta_2 = \beta_1 \exp[i\theta_{12}] P_t/2$ allows $S$ to be rewritten as
\begin{equation}\label{shearapprox}
S = \int \limits_{-\infty}^{\infty}\frac{dk}{2\pi}4k^2e^{2|k|}|\beta_1|^2\left\{  \mathrm{Im}(\omega_1^*)\left(1 - \frac{P_t^2}{4}\right)
+ \frac{P_t}{2}\mathrm{Im}\left[\omega_1^*(2i\sin(\theta_{12}))\right]\right\}.
\end{equation}
Due to the form of $\omega_1$ [see Fig.~\ref{dispersion_plot}], the first term is only nonzero for $|k| \lesssim 0.64$, and the second term is only nonzero for $|k| \gtrsim 0.64$. It is clear that $P_t \sim 1$ reduces the magnitude of the first term, while the contribution of the second term to $S$ depends significantly on the cross-phase $\theta_{12}$ between the eigenmodes.

Having shown that momentum transport can be affected by stable mode activity, we next summarize the main findings of this paper.

\section{Conclusion}
Shear-flow instabilities are widely studied due to their potential to drive turbulence in systems where the turbulent transport of momentum, particles, and heat are of interest. While the linear regime of these instabilities are generally well-understood, saturation and the resulting nonlinear flows are difficult to model. We have presented a nonlinear analysis of an unstable shear layer with piecewise-linear shear flow, showing that the complex conjugate stable linear eigenmode is excited nonlinearly and strongly affects saturation. This result is significant because it represents the first demonstration that nonlinear excitation of linearly stable modes is an important aspect of saturation in global-scale unstable plasma and hydrodynamic systems. Previous studies were limited to quasihomogeneous systems on gyroradius scales\cite{Terry2006,Makwana}. A critical aspect of this work is the development of a mapping technique that allows analytical saturation analyses derived for spatially homogeneous systems to be applied to the strongly inhomogeneous situation of shear flow instability.

Assuming the flow is a linear combination of the linear eigenmodes allows the global state of the system to be described in terms of its behavior at the edges of the shear layer (as is also done to determine the dispersion relation). The nonlinearity, originally written in terms of flow components and their spatial derivatives, is then written in terms of the eigenmodes to demonstrate that unstable modes nonlinearly pump stable modes. This allows the eigenfunctions of this system to be treated similarly to the eigenvectors of previous systems. Using a parameter that quantifies the threshold for a stable mode to impact saturation, we have estimated the impact of stable modes on instability saturation and found it to be significant.

Analysis of the flow associated with stable modes indicates that, at the predicted saturation levels, the fluctuating flow undergoes significant topological changes relative to flows characterized by the unstable mode alone. Such changes may affect the propensity for the turbulent flow structure to generate secondary structures through transient amplification and other processes. Because the system described here is inviscid, this work indicates that stable modes have the potential to modify the evolution of instabilities even when they are not subject to dissipation.

Finally, we consider the contribution of stable modes to momentum transport and give an estimate in terms of the threshold parameter, demonstrating that stable modes can significantly reduce the broadening of the shear layer, thereby counteracting the effect of the unstable modes. One may similarly expect that stable modes can affect other transport channels such as matter entrainment and heat transport. This line of inquiry will be left for future investigations.

\begin{acknowledgments}
The authors would like to thank F.~Waleffe for valuable discussions and insights. Partial support for this work was provided by the Wisconsin Alumni Research Foundation and the U S Department of Energy, Office of Science, Fusion Energy Sciences, under award No.~DE-FG02-89ER53291.
\end{acknowledgments}

\appendix
\section{Coupling Coefficients}
In Eqs.~\eqref{unstablemode} and \eqref{stablemode}, the nonlinear coupling coefficients $C_j,D_j$, which are obtained by expressing the nonlinearities of Eq.~\eqref{NLeigenmode} in terms of the eigenmode amplitudes $\beta_j$, are as follows:
\begin{equation}
\begin{split}
C_1 &= \alpha \left[ ( b_2b_1' + b_1'' ) e^{2|k''|} + ( b_2b_1'' + b_1' ) e^{2|k'|} \right]\\
C_2 &= \alpha \left[ ( b_2b_1' + b_2'' ) e^{2|k''|} + ( b_2b_2'' + b_1' ) e^{2|k'|} \right]\\
C_3 &= \alpha \left[ ( b_2b_2' + b_1'' ) e^{2|k''|} + ( b_2b_1'' + b_2' ) e^{2|k'|} \right]\\
C_4 &= \alpha \left[ ( b_2b_2' + b_2'' ) e^{2|k''|} + ( b_2b_2'' + b_2' ) e^{2|k'|} \right]\\
D_1 &= -\alpha \left[ ( b_1b_1' + b_1'' ) e^{2|k''|} + ( b_1b_1'' + b_1' ) e^{2|k'|} \right]\\
D_2 &= -\alpha \left[ ( b_1b_1' + b_2'' ) e^{2|k''|} + ( b_1b_2'' + b_1' ) e^{2|k'|} \right]\\
D_3 &= -\alpha \left[ ( b_1b_2' + b_1'' ) e^{2|k''|} + ( b_1b_1'' + b_2' ) e^{2|k'|} \right]\\
D_4 &= -\alpha \left[ ( b_1b_2' + b_2'' ) e^{2|k''|} + ( b_1b_2'' + b_2' ) e^{2|k'|} \right],
\end{split}
\end{equation}
where
\begin{equation*}
\alpha = \frac{ik|k'||k''|e^{-|k|-|k'|-|k''|}}{2|k|(b_1-b_2)},
\end{equation*}
with $b_j' \equiv b_j(k')$ and $b_j'' \equiv b_j(k'')$. For convenience, the definition of $b_j(k)$ is repeated here:
\begin{equation*}
b_j = e^{2|k|}\frac{2|k|(\omega_j+k)-k}{k}.
\end{equation*}
Notice that $\alpha(k,k') = \alpha(k,k-k')$ and $C_3(k,k') = C_2(k,k-k')$. Thus, changing the integration variable to $k'' = k-k'$ in the $C_3$ integral yields
\begin{equation*}
\begin{split}
\int_{-\infty}^{\infty}\frac{dk'}{2\pi}C_3(k,k')\beta_1(k'')\beta_2(k') &= \int_{-\infty}^{\infty}\frac{dk'}{2\pi}C_2(k,k'')\beta_1(k'')\beta_2(k')\\
&= \int_{-\infty}^{\infty}\frac{dk''}{2\pi}C_2(k,k'')\beta_1(k'')\beta_2(k-k'')\\
&= \int_{-\infty}^{\infty}\frac{dk'}{2\pi}C_2(k,k')\beta_1(k')\beta_2(k-k'),
\end{split}
\end{equation*}
so the $C_3$ and $C_2$ integrals in Eq.~\eqref{unstablemode} are identical.

\end{document}